\documentclass[prd,aps,twocolumn,showpacs,preprintnumbers,amsmath,amssymb]{revtex4}
\usepackage[dvips]{graphicx}
\usepackage{epsf}
\usepackage{amsmath}
\usepackage{amssymb}

\usepackage{graphicx}
\usepackage{dcolumn}
\usepackage{bm}
\def\be{\begin{equation}}
\def\ee{\end{equation}}
\def\bea{\begin{eqnarray}}
\def\eea{\end{eqnarray}}

\newcommand{\comment}[1]{}
\newcommand{\hphi}{\hat{\phi}}
\newcommand{\tphi}{\tilde{\phi}}

\begin{document}


\date{\today}

\title{A Nonsingular Cosmology with a Scale-Invariant Spectrum of
Cosmological Perturbations from Lee-Wick Theory}

\author{Yi-Fu Cai$^{1}$, Taotao Qiu$^{1}$, Robert Brandenberger$^{1,2,3}$ and Xinmin Zhang$^{1,3}$}

\affiliation{1) Institute of High Energy Physics, Chinese Academy
of Sciences, P.O. Box 918-4, Beijing 100049, P.R. China}

\affiliation{2) Department of Physics, McGill University,
Montr\'eal, QC, H3A 2T8, Canada}

\affiliation{3) Theoretical Physics Center for Science Facilities
(TPCSF), Chinese Academy of Science, P.R. China}


\begin{abstract}

We study the cosmology of a Lee-Wick type scalar field theory.
First, we consider homogeneous and isotropic background solutions
and find that they are nonsingular, leading to cosmological
bounces. Next, we analyze the spectrum of cosmological
perturbations which result from this model. Unless either the
potential of the Lee-Wick theory or the initial conditions are
finely tuned, it is impossible to obtain background solutions
which have a sufficiently long period of inflation after the
bounce. More interestingly, however, we find that in the generic
non-inflationary bouncing cosmology, perturbations created from
quantum vacuum fluctuations in the contracting phase have the
correct form to lead to a scale-invariant spectrum of metric
inhomogeneities in the expanding phase. Since the background is
non-singular, the evolution of the fluctuations is defined
unambiguously through the bounce. We also analyze the evolution of
fluctuations which emerge from thermal initial conditions in the
contracting phase. The spectrum of gravitational waves stemming
for quantum vacuum fluctuations in the contracting phase is also
scale-invariant, and the tensor to scalar ratio is not suppressed.

\end{abstract}

\maketitle

\newcommand{\eq}[2]{\begin{equation}\label{#1}{#2}\end{equation}}

\section{Introduction}

Recently, ideas originally due to Lee and Wick \cite{Lee} were
used to propose \cite{GCW} a ``Lee-Wick Standard Model", a
modification of the Standard Model of particle physics in which
the Higgs mass is stabilized against quadratically divergent
radiative corrections and which in this sense is an alternative to
supersymmetry for solving the hierarchy problem. The Lagrangian
includes new higher derivative operators for each field. These
operators can be eliminated by introducing a set of auxiliary
fields, one for each field of the original model. The higher
derivative terms have opposite sign for both the kinetic and mass
terms, which indicates how the quadratic divergences in the Higgs
mass can be cancelled.

Fields with opposite sign of the kinetic term in the action have
recently been invoked in cosmology to provide models for dark
energy. Fields with negative kinetic energy but positive potential
energy are called ``phantom field" \cite{phantom} and were
introduced to provide a possible mechanism for obtaining an
equation of state of dark energy with an equation of state
parameter $w < -1$, where $w = p / \rho$, $p$ and $\rho$ being
pressure and energy density, respectively. In addition to the
conceptual problems of having phantom fields (see e.g
\cite{Cline}), phantom dark energy models lead to future
singularities. To avoid these problems, the ``quintom model"
\cite{quintom} was introduced. This model contains two scalar
fields, one of them with a regular sign kinetic term, the second
with an opposite sign kinetic term. This model allows for a
crossing of the ``phantom divide", i.e. a transition of the
equation of state from $w < -1$ to $w > -1$. When applied to early
universe cosmology, quintom models can lead to nonsingular
cosmological backgrounds which correspond to a bouncing universe
\cite{quintombounce} \footnote{The use of a second scalar field to
obtain a nonsingular bounce was already discussed in
\cite{Fabio,Peter1}}.

The Higgs sector of the Lee-Wick Standard Model has similarities
with the Lagrangian of a quintom model: the Higgs field has
regular sign kinetic term but the auxiliary field has a negative
sign kinetic term. Thus, it is logical to expect that the
Lee-Wick model might give rise to a cosmological bounce and thus
solve the cosmological singularity problem, in addition to solving
the hierarchy problem. In this paper we show that this expectation
is indeed realized.

Given that the Lee-Wick model leads to a cosmological bounce, the
cosmology of the very early universe may be very different from
what is obtained by studying the cosmology of the Standard Model.
It is possible to introduce a potential for the scalar field in
order to obtain a sufficiently long period of inflation after the
bounce in order to solve the problems of Standard Big Bang
Cosmology and to obtain a spectrum of nearly scale-invariant
cosmological fluctuations. However, this requires fine-tuning of
the potential. On the other hand, given a bouncing cosmology it is
possible that the cosmological fluctuations originate in the
contracting phase, as in the Pre-Big-Bang \cite{PBB} or Ekpyrotic
\cite{Ekp} scenarios. In this paper, we study the generation and
evolution of fluctuations in our Lee-Wick type model. We consider
both vacuum and thermal initial conditions for the fluctuations in
the contracting phase and follow the perturbations through the
bounce, a process which can be done unambiguously since the bounce
is non-singular.

We find that initial quantum vacuum fluctuations in the
contracting phase have the right spectrum to develop into a
scale-invariant spectrum in the expanding phase. What is
responsible for this result is the fact that there is a coupling
of the growing mode in the contracting phase to the dominant
(constant in time) mode in the expanding phase, and that this
coupling scales with co-moving wave-number as $k^2$. The Lee-Wick
model thus leads to a concrete realization of the proposal of
\cite{FB2}  (see also \cite{Wands,Fabio,Peter1} and more recently
\cite{Pinto}) to obtain a scale-invariant spectrum  of
fluctuations from a matter-dominated contracting phase (see also
\cite{Starob} for an analysis of gravitational wave evolution in
this background).

The outline of this paper is as follows. In the following section
we introduce the Lee-Wick scalar field model which we will study
in the rest of the paper. In Section 3 we study the background
solutions of this model, taking initial conditions in the
contracting phase. We show that, at least at the level of
homogeneous and isotropic cosmology, it is easy to obtain a
bouncing cosmology. In Section 4 we study how cosmological
fluctuations  set up in the initial contracting phase pass through
the bounce. The evolution of the fluctuations is well behaved,
Section 5 contains the computation of the spectrum of
gravitational waves, starting from quantum vacuum fluctuations in
the contracting phase. We end with a discussion of our results.

\section{A Lee-Wick Scalar Field Model}

We will take our starting Lagrangian for the scalar field
${\hat{\phi}}$ to be
\be
{\cal L} \, = \, \frac{1}{2} \partial_{\mu} \hphi \partial^{\mu} \hphi - \frac{1}{2M^2}(\partial^2 \hphi)^2
- \frac{1}{2} m^2 \hphi^2 - V(\hphi) \, ,
\ee
where $m$ is the mass of the scalar field and $V$ is its
interaction potential. The second term on the right-hand side is
the higher derivative term, involving a new mass scale $M$.

As discussed in \cite{GCW}, by introducing an auxiliary field
$\tphi$ and redefining the ``normal" scalar field as
\be
\phi \, = \, \hphi + \tphi \, ,
\ee
the Lagrangian takes the form
\be
{\cal L}  =  {1 \over 2} \partial_{\mu} \phi \partial^{\mu} \phi
- \frac{1}{2} \partial_{\mu} \tphi \partial^{\mu} \tphi + \frac{1}{2} M^2 \tphi^2
- \frac{1}{2} m^2 (\phi - \tphi)^2 - V(\phi - \tphi) \, .
\ee
We thus see that $M$ is the mass of the new scalar degree of
freedom, the ``Lee-Wick scalar" which comes from the extra degrees
of freedom of the higher derivative theory. Note that both the
kinetic term and the mass term of the Lee-Wick scalar have the
opposite sign compared the signs for a regular scalar field.   One
may worry that the theory is unstable because of the wrong sign of
the kinetic term of the Lee-Wick scalar \cite{BG,Nak,Gleeson}.
However, as was argued in \cite{Cut}, the perturbative expansion
can be defined in a consistent way and the theory is unitary.
Building on these workds, a recent study shows that Lee-Wick
electrodynamics can be defined consistently as a ghost-free,
unitary and Lorentz invariant theory \cite{Tonder}.

By rotating the field basis, the mass term can be diagonalized.
However, the coupling between the two fields in the interaction
term remains. To be specific, we consider a quartic interaction
term. Thus, the Lagrangian we study is
\be \label{L3}
{\cal L}  =   {1 \over 2}  \partial_{\mu} \phi \partial^{\mu} \phi
- \frac{1}{2} \partial_{\mu} \tphi \partial^{\mu} \tphi + \frac{1}{2} M^2 \tphi^2
- \frac{1}{2} m^2 \phi^2 - \frac{\lambda}{4} (\phi - \tphi)^4 \, .
\ee

\section{Background Cosmology}

In this section we study the background cosmological equations
which follow from coupling the matter Lagrangian (\ref{L3}) to
Einstein gravity. For a homogeneous, isotropic and spatially flat
universe the metric of space-time is
\be
ds^2 \, = \, dt^2 - a(t)^2 d{\bf x}^2 \, ,
\ee
where $t$ is physical time, and ${\bf x}$ denote the co-moving
spatial coordinates. The system of equations of motion consists of
the Klein-Gordon equations
\bea \label{KGeq}
{\ddot \phi} + 3 H {\dot \phi} + m^2 \phi \, &=& \, - \lambda (\phi - \tphi)^3 \nonumber \\
{\ddot \tphi} + 3 H {\dot \tphi} + M^2 \tphi \, &=& \, - \lambda (\phi - \tphi)^3
\eea
for the two scalar fields and the Einstein expansion equation
\be \label{Heq}
H^2 \, = \, \frac{8 \pi G}{3} \bigl[ \frac{1}{2} {\dot \phi}^2 - \frac{1}{2} {\dot \tphi}^2
+ \frac{1}{2} m^2 \phi^2 - \frac{1}{2} M^2 \tphi^2 + \frac{\lambda}{4} (\phi - \tphi)^4 \bigr] \, ,
\ee
where $H = {\dot a}/a$ is the Hubble expansion rate and $G$ is
Newton's gravitational constant. An overdot denotes the derivative
with respect to $t$. Combining these equations leads to the
following expression for the change in the Hubble expansion rate
\be \label{dotHeq}
{\dot H} \, = \,  - 4 \pi G \bigl({\dot \phi}^2 - {\dot \tphi}^2 \bigr) \, .
\ee
from which we immediately see that it is possible for the
background cosmology to cross the ``phantom divide" ${\dot H} =
0$.

Let us take a first look to how it is possible to obtain a
bouncing cosmology in our model. We assume that the universe
starts in a contracting phase and that the contribution of $\phi$
in the equations of motion dominates over that of the Lee-Wick
scalar. This will typically be the case at low energy densities
and curvatures. As the universe contracts and the energy density
increases, the relative importance of $\tphi$ compared to $\phi$
will grow. From (\ref{Heq}) it follows that there will be a time
when $H = 0$ - this is a necessary condition for the bounce point.
From (\ref{dotHeq}) it follows that at the bounce point ${\dot H}
> 0$. Hence, we indeed have a transition from a contracting phase
to an expanding phase, i.e. a cosmological bounce.

Let us now consider the above argument in a bit more detail. For
the moment we will set the interaction Lagrangian to zero, i.e. we
will assume $\lambda = 0$. We begin the evolution during the
contracting phase when the energy density is sufficiently low so
that we expect the contribution of the Lee-Wick scalar to the
total energy density to be small. For these initial conditions,
both matter fields will be oscillating, and the equation of state
will hence be that of a matter dominated universe. In fact, as
follows from the Klein-Gordon equations (\ref{KGeq}) which in this
case reduce to
\bea
{\ddot \phi} + 3 H {\dot \phi} + m^2 \phi \, &=& \, 0 \nonumber \\
{\ddot \tphi} + 3 H {\dot \tphi} + M^2 \tphi \, &=& \, 0
\eea
both scalar fields will be performing oscillations with amplitudes
${\cal A}(t)$ and ${\cal {\tilde A}}(t)$ which are blue-shifting
(i.e. increasing) at the same rate
\be \label{scaling}
{\cal A}(t) \, \sim \, {\cal {\tilde A}}(t) \, \sim \, a(t)^{-3/2} \, .
\ee

Eventually, the oscillations of the field $\phi$ will freeze out.
From studies of chaotic inflation \cite{chaotic} it is well know
that this happens when the amplitude ${\cal A}$ becomes of the
order of the Planck mass $m_{pl}$, more specifically when
\be \label{freeze}
{\cal A} \, = \, (12 \pi)^{-1/2} m_{pl} \, .
\ee
After freeze-out, $\phi$ will slowly roll up the potential and the
equation of state will shift from $w = 0$ to $w \simeq -1$  (where
$w = p / \rho$, $p$ and $\rho$ denoting pressure and energy
density, respectively) leading to a deflationary phase during
which the scale factor is decreasing almost exponentially. This
phase is the time reversal of a period of slow-roll inflation.
However, during this period the Lee-Wick field $\tphi$ is still
oscillating with rapidly increasing amplitude. Hence, its
contribution to the energy density will rapidly catch up to that
of $\phi$.

Let us give a rough estimate of the duration of the deflationary
phase. It will depend crucially on the initial ratio of the energy
density of the Lee-Wick scalar $\tphi$ to that of the regular
scalar $\phi$. Let us denote this ratio by ${\cal F}$. In the
absence of coupling between the two scalar fields, i.e. for
$\lambda = 0$, the ratio will be unchanged during the period of
matter domination when both fields are oscillating. However, once
$\phi$ enters the slow-rolling phase, the amplitude ${\cal{\tilde
A}}$ will increase exponentially according to (\ref{scaling})
while that of $\phi$ will remain virtually unchanged. Thus, the
condition on the duration $\Delta t$ of the deflationary phase is
\be
|H| \Delta t \, \equiv \, N \, = \, \frac{1}{3} log({|{\cal F}|^{-1}}) \, .
\ee
Thus, to obtain a deflationary phase with $N > 50$ (which in the
expanding phase will correspond to a period of inflation of
sufficient length to solve the cosmological problems of the
Standard Big Bang Model) required severe fine-tuning of the
initial conditions. As we will discuss below, this problem may be
even worse if coupling between the two fields is allowed.

Once the contribution of the Lee-Wick scalar to the energy density
catches up to that of the original scalar field, the deflationary
phase will end and a cosmological bounce will occur. Note that
once $H = 0$, the Lee-Wick scalar is still oscillating whereas
$\phi$ is slowly rolling. Thus, ${\dot H} > 0$ and we indeed have
a transition from contraction to expansion. This is a behaviour
which is not possible for Einstein gravity coupled to matter
satisfying the weak energy condition. However, due to the negative
sign of the kinetic term in the Lagrangian, the weak energy
condition is violated in our model. Note that in bouncing
cosmologies obtained in higher derivative gravity models such as
\cite{noghost}, it is the higher derivative gravitational terms
which, when interpreted as matter, lead to a violation of the weak
energy condition.

The duration of the bounce can be estimated as follows: The
maximal amplitude $H_m$ of $|H|$ before and after the bounce is
set by
\be
H_m \, \sim \, m \, ,
\ee
since it is determined by the potential energy at the field value
where the slow rolling of $\phi$ begins. The amplitude of
$\dot{H}$ at the bounce, denoted by ${\dot H}_b$, can in turn be
estimated by
\be
{\dot H}_b \, \sim \, 4 \pi G {\dot{\tphi}}^2 \, \sim \, 4 \pi G m^2 m_{pl}^2
\, \sim \, m^2 \, ,
\ee
where in the first step we have used the fact that the kinetic
energy of $\phi$ is negligible at the bounce, and in the second
step the fact that the bounce is determined by having the same
absolute value of energy densities of $\phi$ and $\tphi$, and that
the field value of $\phi$ at the bounce is about $m_{pl}$. The
bounce time $\Delta t_b$ can now be determined via
\be
{\dot H}_b \Delta t_b \, = \, 2 H_m \, .
\ee
This gives
\be
\Delta t_b \, \sim \, m^{-1} \, .
\ee

Note from the above that the value of $H_m$ is set by the mass of
$\phi$, not the mass of $\tphi$. Similarly, the bounce time is
determined by the mass of the original scalar field and not of its
Lee-Wick partner.

After the bounce, the amplitude of the oscillations of the
Lee-Wick scalar exponentially decreases while $\phi$ is now slowly
rolling down the potential. This is a phase of inflation which is
time-symmetric to the phase of deflation before the bounce. As we
have seen, without fine-tuning of the initial contribution of the
Lee-Wick scalar to the energy density, the period of inflation
will be too small for inflation to solve the various problems of
standard cosmology which inflation was invented to solve
\cite{Guth} (see also \cite{Sato}) (such as the horizon and
flatness problems).

Let us add some comments on the effects of allowing a coupling
between the two scalar fields. We expect that this will lead to a
gradual flow of energy between the regular scalar and the Lee-Wick
scalar such that at an energy density corresponding to the scale
of the Lee-Wick scalar, the energy density in the Lee-Wick scalar
will begin to dominate. Thus, allowing for $\lambda \neq 0$ will
lead to a shorter deflationary phase and may completely eliminate
the period of deflation. Complete elimination of the deflationary
phase will occur if the energy density in $\phi$ is larger than
$M^4$ at $\phi = (12 \pi)^{-1/2} m_{pl}$, where $G \equiv
m_{pl}^{-2}$. This is the case if (making use of (\ref{freeze})
\be
M \, < \, \bigl( (12 \pi)^{-1/2} m m_{pl} \bigr)^{1/2} \, .
\ee

The approximate analytical analysis summarized above is supported
by exact numerical results. We have solved the coupled equations
of motion for the scale factor and the two scalar fields $\phi$
and $\tphi$ numerically. Figure 1 presents the results in the base
of a non-interacting model. We plot the time evolution of the
scalar field $\phi$ (denoted $\phi_1$ in the figure), its Lee-Wick
partner $\tphi$ (denoted by $\phi_2$) and the equation of state
parameter $w$. As is evident, there is a non-singular cosmological
bounce, there is no deflationary phase, but the equation of state
parameter $w$ crosses the phantom divide around the bounce point.
Note that the scalar field $\phi$ stops oscillating near the
bounce, whereas the Lee-Wick scalar continues to oscillate and
therefore increases in magnitude by a large factor during the
latter stages of the contracting phase (which is why we have
plotted the time evolution of $\tphi$ on two different scales).

\begin{widetext}

\begin{figure}[htbp]
\includegraphics[scale=1.1]{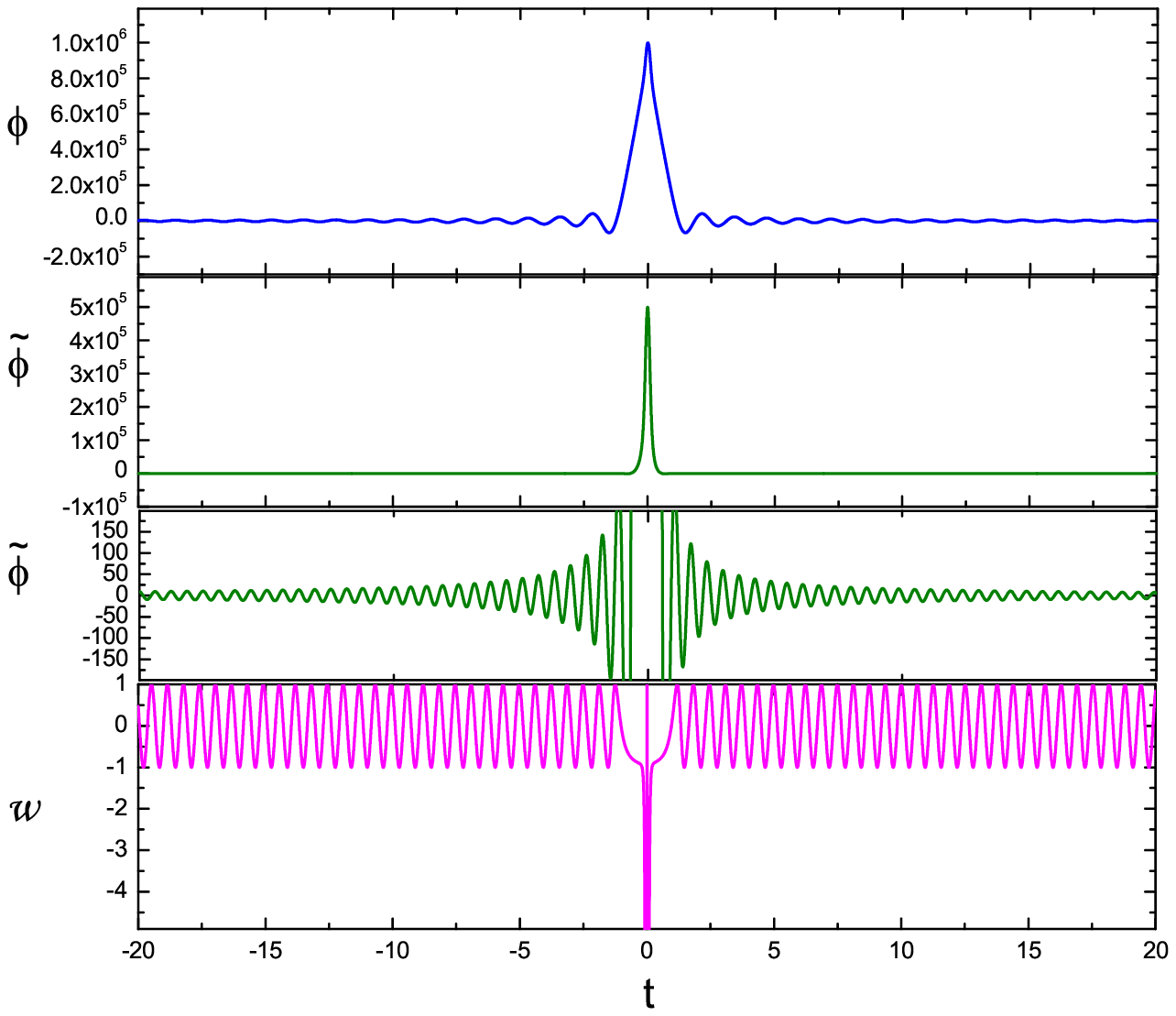}
\caption{Evolution of the background fields $\phi$, $\tphi$ and of
the background equation of state parameter $w$ in a
non-interacting model as a function of cosmic time (horizontal
axis). The background fields are plotted in dimensionless units by
normalizing by the mass $M_{rec} = 10^{-6} m_{pl}$. Similarly, the
time axis is displayed in units of $M_{rec}^{-1}$. The mass
parameters $m$ and $M$ were chosen to be $m = 5 M_{rec}$ and $M =
10 M_{rec}$. The initial conditions were $\phi_i= 1.74 \times
10^3M_{rec}, \,\, \dot\phi_i = 1.44 \times 10^4M_{rec}^2, \,\,
\tilde\phi_i = 8.98 M_{rec}, \,\, \dot{\tilde\phi}_i = -14.08
M_{rec}^2$.}
\end{figure}

\end{widetext}

Note that in the non-interacting model, the bounce is symmetric.
In Figure 2 we present the corresponding figure in the case of an
interacting model with the value of $\lambda$ chosen to be
$\lambda = 1.64 \times 10^{-15}$. In this case, the bounce is
clearly asymmetric. As a second major difference compared to the
simulation of Figure 1, the ratio of masses was chosen to be
almost 100 in this case as opposed to only 2 in the first
simulation. Because of the large ratio of the masses (and the
corresponding initial conditions for which the energy in $\phi$
greatly dominates over than in $\tphi$), the background evolution
enters a brief deflationary phase at the end of the contraction
phase. However, due to the presence of interactions, the energy
density in $\phi$ does not come to dominate again right after the
bounce and hence the period of inflation which would be the time
reversal of the phase of delation is absent.

\begin{widetext}

\begin{figure}[htbp]
\includegraphics[scale=1.1]{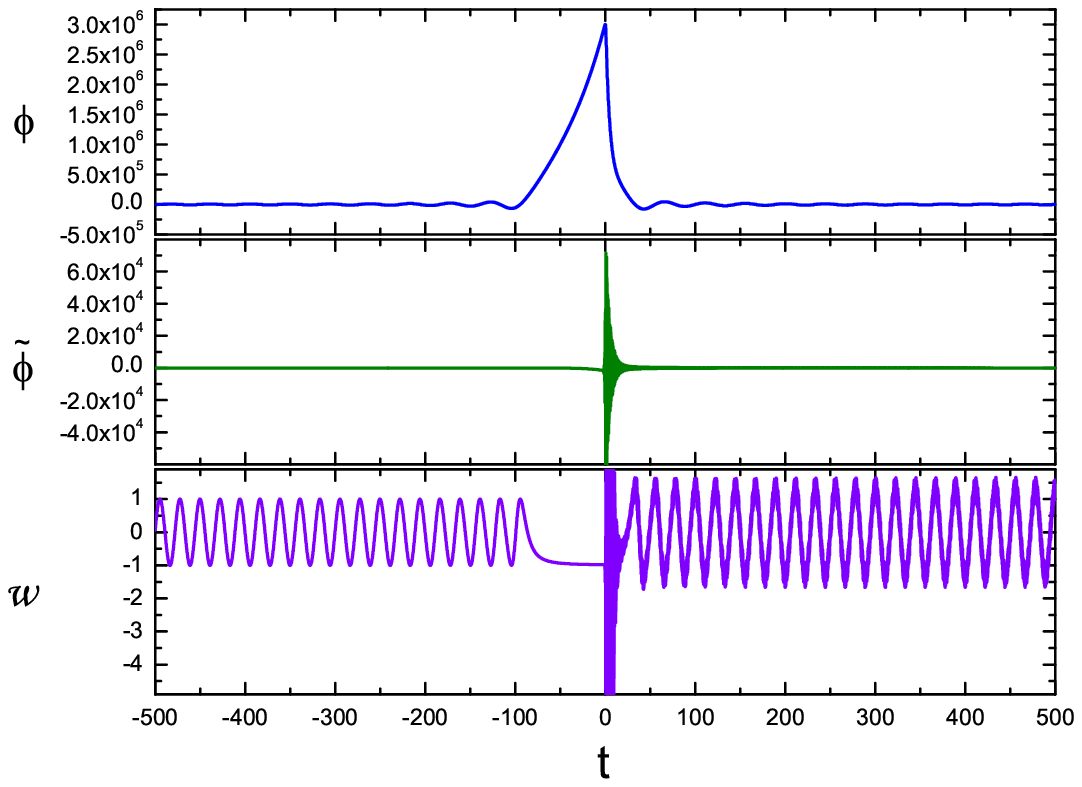}
\caption{Evolution of the background fields $\phi$, $\tphi$ and of
the background equation of state parameter $w$ in a
non-interacting model as a function of cosmic time (horizontal
axis). The background fields are plotted in dimensionless units by
normalizing by the mass $M_{rec} = 10^{-6} m_{pl}$. Similarly, the
time axis is displayed in units of $M_{rec}^{-1}$. The mass
parameters $m$ and $M$ were chosen to be $m = 1.4 \times 10^{-1}
M_{rec}$ and $M = 10 M_{rec}$. The initial conditions were
$\phi_i= - 3.57 \times 10^3 M_{rec}, \,\, \dot\phi_i = 5.56 \times
10^2 M_{rec}^2, \,\, \tilde\phi_i = 2.98 \times 10_{-6} M_{rec},
\,\, \dot{\tilde\phi}_i = -1.39 \times 10^{-6} M_{rec}^2$.}
\label{bgfig}
\end{figure}

\end{widetext}

Finally, in Figure 3 we plot the number of e-foldings of the
deflationary phase as a function of the ratio of $\rho_{\phi}$ to
$\rho_{\tphi}$, in the model without interactions between the two
scalar fields. As predicted by our analytical approximations, the
scaling of the period of deflation as a function of the ratio is
roughly logarithmic.

\begin{figure}[htbp]
\includegraphics[scale=0.8]{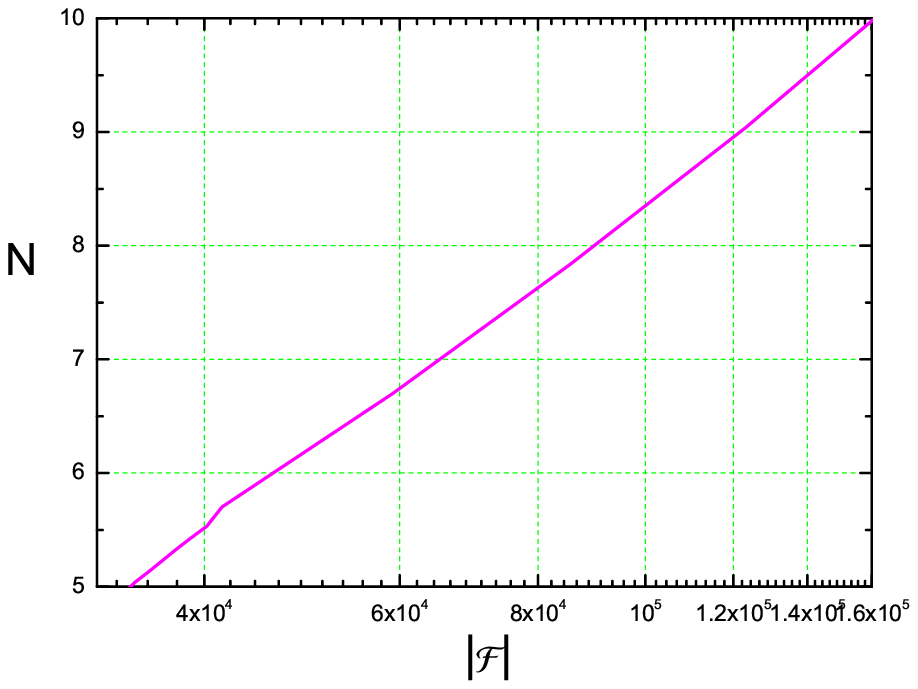}
\caption{Plot of the duration of the deflationary phase as a
function of the ratio of energy densities of $\phi$ and $\tphi$
(horizontal axis). The duration (vertical axis) is shown in terms
of the e-folding number of deflation.}
\end{figure}

\section{Cosmological Fluctuations}

 \subsection{General considerations}

\begin{figure}[htbp]
\includegraphics[scale=0.3]{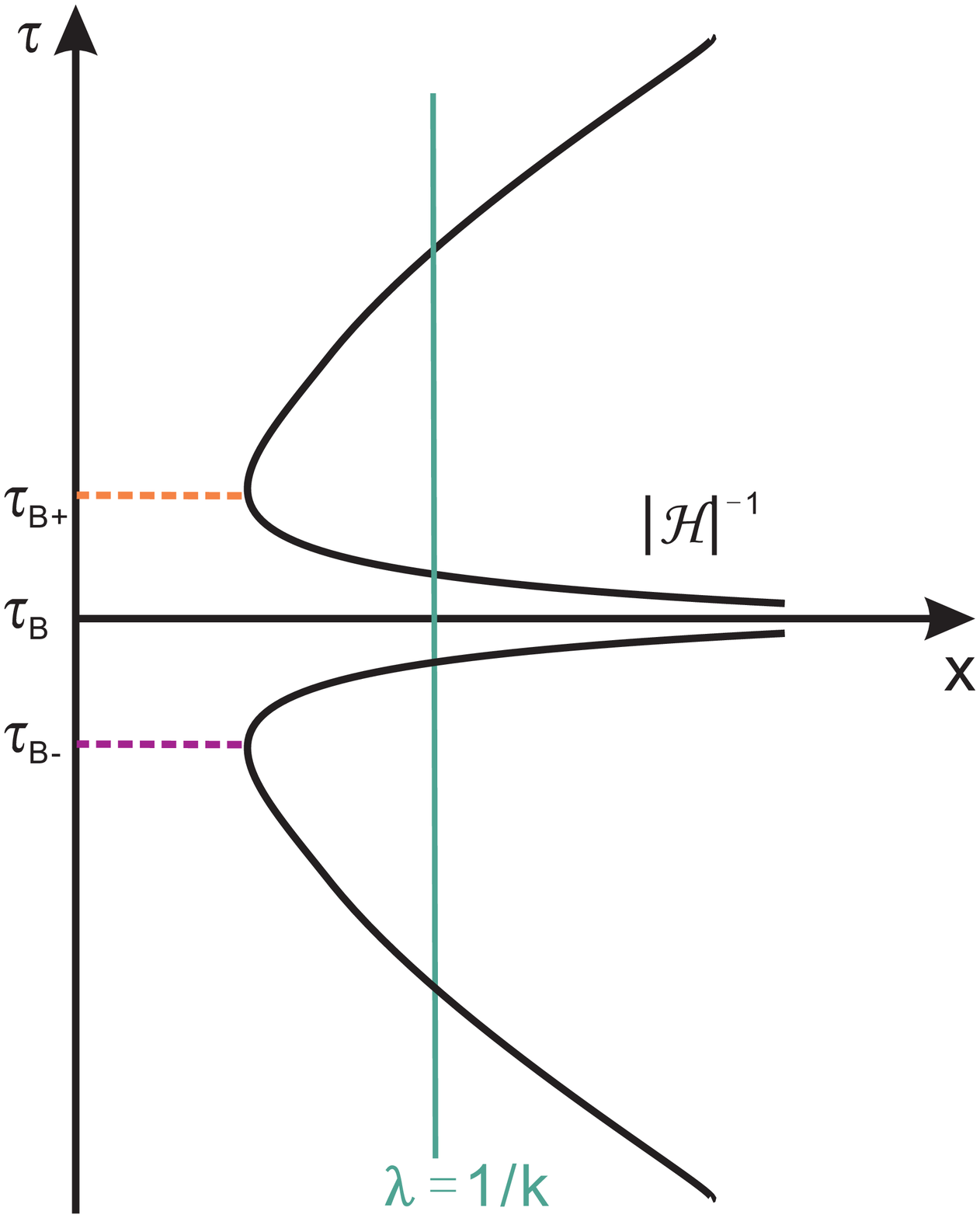}
\caption{A sketch of the evolution of scales in a bouncing
universe. The horizontal axis is co-moving spatial coordinate, the
vertical axis is conformal time. Plotted are the Hubble radius
$|\cal{H}|^{-1}$ and the wavelength $\lambda$ of a fluctuations
with comoving wavenumber $k$. } \label{Fig-st}
\end{figure}

It is useful to first consider the space-time sketch
(\ref{Fig-st}) of our non-singular bouncing cosmology. We choose
the bounce time to correspond to $t = 0$. Long before the bounce,
the equation of state is that of matter. During this period, the
Hubble radius is decreasing linearly and $\dot{H} < 1$.  At a time
denoted $- t_R$ (in analogy with the notation in inflationary
cosmology) there is a transition to a period of deflation during
which the Hubble radius $|H|^{-1}$ is constant. However, as argued
in the previous section, this period will be of short duration and
ends at a time $- t_i$ when a brief  bouncing phase covering the
time interval $[- t_i < t < t_i$ begins. During this period
$\dot{H} > 0$. After the bouncing phase there is a short period of
inflation lasting from $t_i$ to $t_R$, after which the universe
enters a matter-dominated expansion phase with $\dot{H} < 0$.

In Figure \ref{Fig-st} we also plot the evolution of the physical
length corresponding to a fixed co-moving scale. This scale is the
wavelength of the fluctuation mode $k$ ($k$ standing for the
co-moving wavenumber) which we want to follow. The wavelength
begins in the matter-dominated phase of contraction on sub-Hubble
scale, exits the Hubble radius during this phase at a time which
we denote $-t_H(k)$, and re-enters the Hubble radius during the
matter-dominated phase at the time $t_H(k)$.

Note that if the energy density at the bounce point is given by
the scale $\eta$ of Grand Unification ($\eta \sim 10^{16} {\rm
GeV}$), then the physical wavelength of a perturbation mode
corresponding to the current Hubble radius is of the order of $1
{\rm mm}$, i.e. in the far infrared. In this sense, the evolution
of fluctuations in this bouncing cosmology is free of the
trans-Planckian problem \cite{RHBrev,Jerome} which effects the
evolution of fluctuations in all inflationary models in which the
period of inflation lasts more than about $70$ e-foldings (this
number assumes that the scale of inflation is of the order of
Grand Unification).

Since our bounce is non-singular, the computation of the evolution
of fluctuations is free of the matching condition ambiguities
which affect the study of fluctuations in singular bouncing
cosmologies such as the Ekpyrotic scenario (see
\cite{Lyth,Khoury,FF1,Hwang,Peter3,Durrer} for some early papers
on the problem of matching fluctuations through the bounce in
Ekpyrotic cosmology).

There is another important difference in the study of cosmological
fluctuations between non-singular bouncing cosmologies and the
inflationary scenario. It is usually argued that the exponential
expansion of space during inflation red-shifts any pre-existing
matter and the related matter fluctuations, leaving behind a
vacuum matter state. Thus, perturbations in this setup are quantum
vacuum fluctuations \footnote{In light of the trans-Planckian
problem for inflationary fluctuations, one may view this
prescription with some scepticism.}. On the other hand, in a
bouncing cosmology the fluctuations are set up at low densities
and temperatures in the contracting phase. There is no mechanism
that red-shifts initial classical fluctuations. Thus, there is no
reason to prefer vacuum over thermal initial perturbations. In the
following, we will consider both choices.

\subsection{Equations for cosmological perturbations}

We begin by writing the metric including cosmological fluctuations
in longitudinal gauge, assuming that there is no anisotropic
stress (see \cite{MFB} for a comprehensive discussion of the
theory of cosmological perturbations and \cite{RHBrev1} for a
briefer survey)
\be
ds^2 \, = \, a(\eta)^2 \bigl[ (1 + 2 \Phi) d\eta^2 - (1 - 2 \Phi) d{\bf x}^2 \bigr]  \, ,
\ee
where $\Phi({\bf x}, t)$ is the generalized Newtonian
gravitational potential which represents the metric fluctuations.
It is convenient to write the equations in terms of conformal time
$\eta$ defined via $dt = a(t) d\eta$.

The Einstein equations linearly expanded in $\Phi$ lead to the
following equation of motion for the Fourier mode of $\Phi$ with
co-moving wave-number $k$:
\bea \label{Phieq}
\Phi^{\prime \prime} &+& 2 \bigl( {\cal H} - \frac{\phi^{\prime \prime}}{\phi^{\prime}} \bigr) {\Phi^{\prime}}
+ 2\bigl( {\cal H}^{\prime} - {\cal H} \frac{\phi^{\prime \prime}}{\phi^{\prime}} \bigr) \Phi
+ k^2 \Phi \nonumber \\
&=& \, 8 \pi G \bigl( 2 {\cal H} + \frac{\phi^{\prime \prime}}{\phi^{\prime}} \bigr) \tphi^{\prime} \delta \tphi
\eea
where the derivative with respect to conformal time is denoted by
a prime, ${\cal H} \equiv a'/a$, and $\delta \tphi$ is the
fluctuation in $\tphi$. In deriving this equation, we have assumed
that the background is dominated by the field $\phi$. This will be
the case except at the bounce.

In inflationary cosmology, it has proven to be convenient to use
the variable $\zeta$, the curvature fluctuation in co-moving
gauge, which in terms of $\Phi$ is given by
\be
\zeta \, = \, \Phi + \frac{{\cal H}}{{\cal H}^2 - {\cal H}^{\prime}} \bigl( \Phi^{\prime} + {\cal H} \Phi \bigr) \, .
\ee
In any eternally expanding universe in which $1 + w \neq 0$, the
variable $\zeta$ is conserved on super-Hubble scales in the
absence of entropy fluctuations \cite{Bardeen,BST,BK}, since -
neglecting terms of the order $k^2$ - the equation of motion
(\ref{Phieq}) for $\Phi$ is equivalent to
\be
(1 + w) {\dot \zeta} \, = \, 0 \, .
\ee

When considering the quantum theory of cosmological perturbations,
it is important to identify the fluctuation variable which has
canonical kinetic term. It is with respect to this variable,
commonly denoted by $v$, that the canonical commutation relations
must be imposed (see \cite{Sasaki,Mukh} for the quantum theory of
cosmological perturbations). It turns out that the variable is
simply related to $\zeta$
\be
v \, = \, z \zeta \, ,
\ee
where the background variable $z$ is the following combination of
the background metric and the background matter field $\phi$ (for
simplicity we are assuming here only one matter field)
\be \label{zdef}
z \, = \, \frac{a \phi^{\prime}}{{\cal H}} \, .
\ee
If the equation of state is constant in time, then $z(\eta)$ is
proportional to $a(\eta)$.

The equation of motion for $v$ is
\be \label{veq}
v^{\prime \prime} + \bigl[ k^2 - \frac{z^{\prime \prime}}{z} \bigr] v \, = \, 0 \, ,
\ee
On sub-Hubble scales, it follows from (\ref{veq}) that $v$ is
performing harmonic oscillations as a function of conformal time.
On the other hand, on super-Hubble scales $v$ is frozen in and
$v(\eta) \sim z(\eta)$.

In terms of the variable $z$, the relationship between the metric
fluctuation $\Phi$ and the canonical field $v$ takes the form
\cite{MFB}
\be \label{rel}
\Phi \, = \,  \frac{4 \pi G}{k^2} \frac{\phi\prime^2}{\cal H} \bigl(\frac{v}{z}\bigr)^{\prime}
\, .
\ee

The variable $\zeta$ has proven to be a convenient variable to use
in inflationary cosmology. It was therefore taken for granted that
it would also be a useful variable in bouncing cosmologies, and
that it would remain conserved between when the mode $k$ exits the
Hubble radius during the period of contraction at the time
$-t_H(k)$ and the time $t_H(k)$ of re-entry in the expanding
phase. In the context of singular bouncing cosmologies, the
Hwang-Vishniac \cite{HV} (Deruelle-Mukhanov \cite{DM}) matching
conditions for fluctuations across the singularity lead to the
conclusion that $\zeta$ should be conserved. However, as pointed
out in \cite{Durrer}, the applicability of these matching
conditions is questionable since the matching conditions are not
satisfied by the background.

Non-singular bouncing cosmologies do not require ad-hoc matching
conditions - the fluctuations can be followed through the bounce
(as long as their amplitude remains sufficiently small such that
linear perturbation theory does not break down). As has recently
been shown in several examples of non-singular bounces, the
equation of motion for $\zeta$ develops singularities around the
bounce point \cite{BFS,ABB,CQBPZ}, whereas the equation of motion
for $\Phi$ remains well defined. One of the reasons for the
singularities in the equation of motion for $\zeta$ is that the
co-moving gauge has a singularity at a cosmological bounce. Thus,
in the following we will follow the evolution of the fluctuations
in terms of $\Phi$.

If the initial fluctuations in the contracting phase are due to
thermal matter, then the initial values of $\Phi$ and its
derivative follow from the perturbations in the energy density of
matter. If, on the other hand, we assume vacuum initial
fluctuations, then the initial inhomogeneities are given in terms
of the canonical variable $v$ and the initial values of $\Phi$ and
${\dot \Phi}$ must be induced from $v$ via the relation
(\ref{rel}).

\subsection{General solutions}

Let us briefly review the general solution of the equation of
motion (\ref{Phieq}) for $\Phi$ on super-Hubble scales. We will
keep the discussion quite general in this subsection and assume
that the equation of state parameter is given by some $w$. In this
case, $a(t)$ scales as
\be
a(t) \, \sim \, t^p
\ee
with
\be
p \, = \, \frac{2}{3(1 + w)} \, .
\ee
From the definition of conformal time $\eta$ it then follows that
\be
\eta \, \sim \, t^{1 - p} \, .
\ee

The condition for the Hubble radius crossing time $t_H(k)$ for a
mode with co-moving wave-number $k$ is
\be
a(t_H(k)) k^{-1} \, = \, H^{-1}(t_H(k)) \, \sim t_H(k) \, .
\ee
Hence
\be
\eta_H(k) \, \sim \, k^{-1} \, .
\ee

As is well known, one of the two modes of $\Phi$ on super-Hubble
scales is constant, whereas the other is decaying in an expanding
universe and growing in a contracting one. Specifically we have
(see e.g. \cite{FB2})
\be
\Phi(k, \eta) \, = \, D(k) + S(k) \eta^{-2 \nu} \, ,
\ee
where
\be
2 \nu \, = \, \frac{5 + 3 w}{1 + 3 w} \, ,
\ee
and where $D(k)$ and $S(k)$ are independent of time and carry the
information about the spectra of the two modes. In the following,
we will determine the spectra of these two modes for various
thermal and vacuum initial conditions.

\subsection{Thermal initial conditions}

Here we assume that the initial fluctuations are given by thermal
matter perturbations. As was done in the case of string gas
cosmology \cite{NBV,BNPV2} (see \cite{RHBrev3} for a recent
comprehensive review), we follow the matter perturbations up to
Hubble radius crossing and then convert to metric fluctuations by
making use of the perturbed Einstein constraint equation (the
time-time component of the perturbed Einstein equations) which
reads
\be \label{pertEE}
- 3 {\cal H} \left( {\cal H} \Phi + \Phi^{'} \right) + \nabla^2 \Phi
\, = \, 4 \pi G a^2 \delta T^0{}_0 \, .
\ee
In the above, $\delta T^0_0$ is the fluctuation in the energy
density, and $\nabla$ is the co-moving spatial gradient. At Hubble
radius crossing all three terms on the left-hand side of the above
equation are of the same order of magnitude. Hence, modulo a
constant of the order $1$, the Fourier space correlation function
of $\Phi$ becomes
\be \label{Phicor}
< | \Phi(k)|^2 > \, = \, 16 \pi^2 G^2 k^{-4} a^4 < |\delta T^0_0(k)|^2 > \, .
\ee
where the pointed brackets indicate ensemble averaging.

The energy density fluctuations are determined by thermodynamics.
First, we express the momentum space energy density correlation
function for co-moving wave-number $k$ in terms of the r.m.s.
position space mass fluctuation
\be \label{mass}
\delta M(R)^2 \, = \, R^3 <|\delta T^0_0(k)|^2> \, ,
\ee
where $R = a k^{-1}$ is the physical radius of the region
corresponding to the wave-number $k$. The mass fluctuations are
determined by the specific heat capacity $C_V$
\be \label{mass2}
\delta M(R)^2 \, = \, T^2 C_V (R) \, ,
\ee
where $T$ is the temperature of the system. For a gas of point
particles, the heat capacity is proportional to $R^3$, i.e.
\be \label{specheat}
C_V (R) \, = \, c_V T^3 R^3 \, ,
\ee
where $c_V$ is a constant. Note that this result is in agreement
with the intuition that on scales larger than the thermal
correlation length $T^{-1}$, the heat capacity scales as a random
walk.

Inserting (\ref{mass}), (\ref{mass2}) and (\ref{specheat}) into
(\ref{Phicor}) we obtain the following power spectrum for $\Phi$
\be
k^3 < | \Phi(k)|^2 > \, = \, 16 \pi^2 G^2 k^{-1} T^5 c_V \, .
\ee
We need to evaluate this expression at Hubble radius crossing
$t_H(k)$ since we will be using the corresponding value as the
initial condition for the evolution of $\Phi$ on super-Hubble
scales:
\be \label{horizon1}
k^3 < | \Phi(k)|^2 > (t_H(k)) \, = \, 16 \pi^2 G^2 k^{-1} T^5(t_H(k)) c_V \, .
\ee

We now use (\ref{horizon1}) to infer the power spectra of the $S$
and $D$ modes in the case of thermal matter initial conditions. We
are assuming that the initial value of $\Phi$ at Hubble radius
crossing gets distributed equally among the two modes The power
spectrum $P_D(k)$ of the constant mode $D$ is the same as that of
$\Phi$ at Hubble radius crossing
\be
P_D(k) \, \sim \, k^{-1 + \frac{5p}{1 - p}} \,
\ee
where the second exponent comes from making use of
\be
T(t_H(k)) \, \sim \, a^{-1}(t_H(k)) \, \sim \, t_H(k)^{-p} \, \sim \, k^{\frac{p}{1 - p}} \, .
\ee

The power spectrum of $S$ is the spectrum of $\Phi$ at Hubble
radius crossing modulated by the factor $\eta_H(k)^{2 \nu}$:
\be
P_S(k) \, \sim \, k^{-1 + \frac{5p}{1 - p} - 4 \nu} \, .
\ee

In the example we are interested in, fluctuations leave the Hubble
radius is the matter epoch and hence $p = 2/3$ and $w = 0$. Thus,
from the above we see that the spectra of $D$ and $S$ scale as
\bea
P_D(k) \, &\sim& \, k^9 \nonumber \\
P_S(k) \, &\sim& \, k^{-1}  \, .
\eea
The spectrum of the $D$ mode is extremely blue. The blue tilt is
due to the thermal suppression of the spectrum at large
wavelengths. It is also easy to understand why the spectrum of the
$S$ mode is less blue than that of the $D$ mode: the $S$ mode
grows on super-Hubble scales, and large wavelength modes
experience the growth for a longer period of time. Our calculation
shows that the difference in growth on super-Hubble scales
dominates over the thermal suppression of long wavelength modes.

\subsection{Vacuum initial conditions}

Vacuum initial conditions are given in terms of the canonically
normalized variable $v_k$ being in its quantum vacuum state
\cite{MFB}
\be
v_k(\eta) \, \sim \, k^{-1/2} e^{i \eta k} \, .
\ee
Inserting this into (\ref{rel}) and making use of the fact that on
sub-Hubble scales the derivative of the oscillating factor
dominates over the derivative of other terms leads to the
following initial conditions in terms of $\Phi$:
\be \label{Phiresult}
\Phi_k(\eta) \, \sim \, i \frac{4 \pi G}{k^{3/2}} \frac{\phi\prime^2}{z \cal{H}} \, ,
\ee
i.e. a spectrum which is proportional to $k^{-3/2}$. The same
conclusion can be reached \cite{Khoury} by starting with vacuum
fluctuations in the matter field $\phi$ and inserting the result
into the equations expressing $\Phi$ and ${\dot{\Phi}}$ in terms
of the matter field. Making use of (\ref{zdef}) to eliminate $z$
and of the background Friedmann equation to eliminate ${\dot
\phi}$ in favor of $H$ we find the following result for the power
spectrum of $\Phi$:
\bea \label{ispectrum}
P_{\Phi}(k, \eta)  & \equiv & \frac{1}{2 \pi^2} k^3 |\Phi_k(\eta)|^2 \\
& \simeq &  \frac{3}{\pi} \bigl(\frac{H(\eta)}{m_{pl}} \bigr)^2 \, . \nonumber
\eea

Making use of the definition of $z$ from (\ref{zdef}) it follows
that the time dependent terms in (\ref{Phiresult}) scale as
$H^{-1}$. Thus,
\be
\Phi_k(t) \, \sim \, k^{-3/2} t^{-1} \, ,
\ee
which allows us to evaluate the result at Hubble radius crossing
\be \label{result}
\Phi_k(t_H(k)) \, \sim \, k^{-\frac{3}{2} + \frac{1}{1 - p}} \, .
\ee

The above result (\ref{result}) allows us to compute the spectra
of both $D$ and $S$ modes of $\Phi$ on super-Hubble scales,
assuming - as we did in the previous subsection - that
$\Phi_k(t_H(k))$ sources both modes equally:
\bea
\Phi_D(k) \, & \sim & \, k^{- \frac{3}{2} + \frac{1}{1 - p}} \, \label{Dresult} \\
\Phi_S(k) \, & \sim & \, k^{- \frac{3}{2} + \frac{1}{1 - p} - 2 \nu} \, .
\eea
In the case we are interested in $p = 2/3$, $w = 0$ and $2 \nu =
5$ we obtain
\bea
\Phi_D(k) \, & \sim & \, k^{3/2} \, \\
\Phi_S(k) \, & \sim & \, k^{-7/2} \, .
\eea
Note that, as pointed out in \cite{FB2}, the $S^{-}$ mode leads to
a scale-invariant spectrum of fluctuations of $\zeta$ in the
contracting phase.

This compares to the results obtained for Ekpyrotic type
contraction \cite{Khoury} where $p = 0$, $w = \infty$ and $2 \nu =
1$ and therefore
\bea
\Phi_D(k) \, & \sim & \, k^{-1/2} \, \\
\Phi_S(k) \, & \sim & \, k^{-3/2} \, ,
\eea
which leads to a scale-invariant spectrum for the $S$ mode which
is growing in the phase of contraction.

As we expect from the Hwang-Vishniac (Deruelle-Mukhanov) matching
conditions, the $S-$ mode will couple with a $k^2$ suppression to
the dominant mode in the expanding phase. If this is realized, we
will obtain a scale-invariant spectrum of curvature fluctuations
in the expanding phase. In the following subsection we will evolve
the fluctuations through the non-singular bounce and infer the
spectrum at late times. We indeed find a late-time scale-invariant
spectrum.

\subsection{Evolution of the fluctuations through the bounce}

Let us step back and write down the equation of motion for $\Phi$
in a slightly modified form (which is equivalent to (\ref{Phieq})
except that we allow for a general speed of sound $c_s$ which is
equal to 1 in our scalar field model)
\be \label{Phieq2}
\Phi_k'' + 2\sigma {\cal H}\Phi_k' +  k^2 c_s^2 \Phi_k \, = \, 0 \, ,
\ee
where
\be
\sigma \, \equiv \, -\frac{\ddot H}{2H\dot H} \, .
\ee

\centerline{Contracting phase}

In the contracting phase the equation (\ref{Phieq2}) takes the
form
\be \label{perteqc}
\Phi_k'' + \frac{1+2\nu_c}{\eta-\tilde\eta_{B-}}\Phi_k' + k^2 c_s^2 \Phi_k \, = \, 0 \, ,
\ee
with
\be
\nu_c \, \equiv \, \frac{5+3w_c}{2(1+3w_c)} \, ,
\ee
where the subscript ``c" indicates that we are discussing the
contracting phase. The general analytical solution is
\bea
\Phi_k \, = \, (\eta-\tilde\eta_{B-})^{-\nu_c} \bigg\{
k^{-\nu_c}D_-J_{\nu_c}[c_sk(\eta-\tilde\eta_{B-})]  \nonumber \\  +
k^{\nu_c}S_-J_{-\nu_c}[c_sk(\eta-\tilde\eta_{B-})] \bigg\}~,
\eea
where the coefficients $D_-$ and $S_-$ can be determined by the
initial condition of the gravitational potential as described in
the two previous subsections for different sets of initial
conditions. In the above, $\eta_{B-}$ is a fixed time that
corresponds to when the singular bounce would occur if the
universe were to remain matter-dominated.

Note that, when the wavelength of the perturbation is larger than
Hubble radius with $k\ll |{\cal H}|$, the asymptotical form of
$\Phi_k$ can be written as
\be \label{asolc}
\Phi^c_k\, = \, \bar D_-+\frac{\bar
S_-}{(\eta-\tilde\eta_{B-})^{2\nu_c}}~,
\ee
where we define
\be
\bar D_- \, \equiv \, \frac{c_s^{\nu_c} D_-}{2^{\nu_c}\Gamma(1+\nu_c)}~,~~
\bar S_- \, \equiv \, \frac{2^{\nu_c} S_-}{c_s^{\nu_c}\Gamma(1-\nu_c)}~,
\ee

As discussed in previous subsections, the $\bar D_-$ mode is
constant and the $\bar S_-$ mode is growing in a contracting
universe. Note from the definition of $\zeta_k$ we have to leading
order in $k$
\be
\zeta^c_k \, = \, \frac{5+3w_c}{3(1+w_c)}\bar D_-~.
\ee
Thus, to this order, in the contracting phase $\zeta_k$ is
determined by the constant mode of $\Phi_k$ which is sub-dominant.
As discussed in detail in \cite{Khoury} the $S$ mode does effect
$\zeta$ when $k^2$ corrections to the solutions are taken into
account.

If we match the asymptotic form for $\Phi$ (\ref{asolc}) to the
initial power spectrum of $\Phi$ (see (\ref{ispectrum})) at Hubble
radius crossing and assume that the initial power is equally
distributed into the two modes, we obtain
\be \label{Sspectrum}
\frac{1}{2 \pi^2} k^3 |S_{-}(k)|^2 \, \simeq \, \frac{3}{\pi} m_{pl}^2 t_H(k)^{-2} \eta_H(k)^{4 \nu_c} \, ,
\ee
where the subscript $H$ stands for the time of Hubble radius
crossing.

\centerline{Bouncing phase}

As we have shown in the section on the background dynamics, the
contribution of the higher derivative terms in the Lee-Wick model
becomes more and more important as the universe contracts and will
lead to a non-singular bounce. Thus, the universe will exit from
the phase of matter-dominated contraction at some time $t_{B-}$,
and then the EoS of the universe will cross $-1$ and fall to
negative infinity rapidly. Correspondingly, the Hubble parameter
reaches zero and leads to a bounce of the universe at the time
$\eta_B$. After the bounce, the Lee-Wick field will recover its
normal state with the higher derivative terms rapidly decreasing
in importance.

It is rather complicated to solve the perturbation equation
directly from Eq. (\ref{Phieq2}). In order to solve the equation
analytically, we need to make some approximations to simplify it.
Our approximation consists of choosing a convenient modelling of
the Hubble parameter near the bounce of the form
\be
H \, = \, \alpha t
\ee
with some positive constant $\alpha$ which has dimensions of $k^2$
and whose magnitude is set by the microphysics of the bounce, in
our case by the mass $M$ of the Lee-Wick scalar. The time of the
bounce was chosen to be $t = 0$. In this case, we can obtain an
analytical form for the comoving Hubble parameter in the bouncing
phase:
\bea
{\cal H} \, &=& \, \frac{\frac{y}{3}(\eta-\eta_B)}{1-\frac{y}{6}(\eta-\eta_B)^2}~, \nonumber \\
y \, &=& \,\frac{12}{\pi}\alpha a_B^2~,
\eea
where $a_B$ denotes the value of the scale factor at the bounce
point $\eta_B$.

Since the above parametrization should be valid only in the
neighborhood of the bounce point, the quadratic and higher order
terms of $|\eta-\eta_B|$ can be neglected. Consequently, the
perturbation equation takes the following form
\be \label{perteqb}
\Phi_k'' + 2y(\eta-\eta_B)\Phi_k' + (c_s^2k^2+\frac{2}{3}y)\Phi_k \, = \, 0~.
\ee
The solution of this equation can be written as
\begin{eqnarray}
\Phi_k \, &=& \ \bigg\{ E_kH_l[\sqrt{y}(\eta-\eta_B)] +
F_k~_1F_1[-\frac{l}{2},\frac{1}{2},y(\eta-\eta_B)^2] \bigg\}
\nonumber\\ && \, \times \exp[-y(\eta-\eta_B)^2]~,
\end{eqnarray}
which is constructed from the $l$-th Hermite polynomial and a
confluent hypergeometric function with
\be
l \, \equiv \, -\frac{2}{3}+\frac{c_s^2k^2}{2y} \,
\ee
and two undetermined coefficients $E_k$, $F_k$. These two
functions are linearly independent, and their asymptotical
behaviors are mainly determined by the parameter $l$.

When $c_s^2k^2 \gg y$, i.e. the wave-number of the mode is larger
than the mass scale of the bounce, then both functions are
oscillating. This case was already studied in Ref. \cite{CQBPZ}.

However, in the current paper we are interested in the opposite
limit, the limit in which the wavelength is much larger than the
inverse mass scale of the bounce, i.e. the limit when
\be \label{largecond}
c_s^2k^2 \, \ll \, y \, .
\ee
As we have argued in the section of our paper on the background
evolution, the bounce takes place very fast and thus the condition
(\ref{largecond}) will be satisfied for all wavelengths we are
interested in. In this case, we can expand the solution of the
perturbation equation in a power series in terms of
$\sqrt{y}(\eta-\eta_B)$. Then the solution is given by
\begin{eqnarray} \label{asolb}
\Phi^b_k \, &=& \, \hat{F}_k + \hat{E}_k\sqrt{y}(\eta-\eta_B) +
(-1-l)\hat{F}_ky(\eta-\eta_B)^2 \nonumber\\ &&
\, + O(y^\frac{3}{2}(\eta-\eta_B)^3)~,
\end{eqnarray}
with
\begin{eqnarray}
\hat{E}_k \, &\equiv& \, -\frac{2^{1+l}\sqrt{\pi}}{\Gamma(-\frac{l}{2})}E_k~,~~ \\
\hat{F}_k \, &\equiv& \, \frac{2^{l}\sqrt{\pi}}{\Gamma(\frac{1-l}{2})}E_k+F_k~,
\end{eqnarray}
and the subscript ``b" represents the bouncing phase. In this
case, we have
\be
\zeta^b_k \, \simeq \, \hat{F}_k[1+\frac{c_s^2k^2}{2}(\eta-\eta_B)^2]~.
\ee
Therefore, the conservation of $\zeta_k$ is realized by the mode
$\hat{F}_k$ when the bounce is fast enough.

Now we study how to establish the coefficients $\hat{E}_k$ and
$\hat{F}_k$. We need to use the Hwang-Vishniac \cite{HV}
(Deruelle-Mukhanov \cite{DM}) matching condition to link the
fluctuations in contracting phase with those in the bouncing phase
at the momentum $\eta_{B-}$. Note that since we are matching two
contracting universes across a non-singular surface, the
background satisfies the matching conditions, unlike the situation
in the Ekpyrotic scenario with a singular bounce, Thus, it is
justified to apply the matching conditions \cite{Durrer}.

The matching conditions say that both $\Phi_k$ and
\be
\hat\zeta_k \,  \equiv \, \zeta_k+\frac{c_s^2k^2}{3}\frac{\Phi_k}{{\cal
H}^2-{\cal H}'}
\ee
are continuous on the matching surface of constant energy density.
Taking use of matching conditions in the solutions (\ref{asolc})
and (\ref{asolb}), we can obtain the following relations:
\begin{eqnarray}\label{match1}
&\hat{E}_k\sqrt{y}(\eta_{B-}-\eta_B) \, = \, -(\frac{1}{3}+2l)\Phi^c_k-\hat\zeta^c_k|_{B-}~, \nonumber\\
&\hat{F}_k \, = \, (\frac{4}{3}+2l)\Phi^c_k+\hat\zeta^c_k|_{B-}~.
\end{eqnarray}
These relations show that the constant and growing modes of
gravitational potential get mixed during the bounce. However, if
we consider large wavelengths compared to the duration of the
bounce, the second relation shows that $\zeta_k$ is indeed
conserved across the bounce.

\centerline{Expanding phase}

After the bounce, the higher derivative terms of Lee-Wick field
rapidly decay. Therefore, a phase of matter-dominated expansion
starts at the time $\eta_{B+}$. In the absence of interactions
between the two scalar fields, the background cosmology will be
time-symmetric about the bounce point. During the period after
$\eta_{B+}$ the background evolution is the time reverse of the
contracting phase. In the case $\lambda \neq 0$ an asymmetric
bounce is possible. To render our analysis more general, we assume
that the equation of state in the expanding phase is $w_e$ which
could be different from that in the contracting stage which is
$w_c$.

The equation of motion for the gravitational potential is similar
as Eq. (\ref{perteqc}) but with the indexes
\be
\nu_e \, \equiv \, \frac{5+3w_e}{2(1+3w_e)}
\ee
and ``$B+$" instead of $\nu_c$ and ``$B-$". Then the solution
on super-Hubble scales take the form
\be
\Phi^e_k \, = \, \bar D_++\frac{\bar S_+}{(\eta-\tilde\eta_{B+})^{2\nu_e}}~,
\ee
with
\be
\tilde\eta_{B+} \, \equiv \, \eta_{B+}-\frac{2}{1+3w_e}\frac{1}{{\cal
H}_{B+}} \, .
\ee
The $\bar D_+$ mode of the gravitational potential is constant in
time, as is the $\bar D_-$ mode in contracting phase. However, the
role of the $S$ mode is very different. In the expanding phase
$\bar S_+$ is the sub-dominant decreasing mode, whereas in the
contracting phase $\bar S_-$ is the dominant expanding mode.
Therefore, the dominant mode of the curvature perturbation in the
period of expansion  is $\bar D_+$. As we will shown in the
following, it inherits contributions from both $\bar D_-$ and
$\bar S_-$ since these modes mix during the bounce.

To determine the coefficients of the two modes in the expanding
phase, we need to apply the matching condition again, this time at
the surface $\eta_{B+}$. A straightforward calculation yields
\begin{eqnarray} \label{match2}
&\hat{E}_k\sqrt{y}(\eta_{B+}-\eta_B) \, = \, -(\frac{1}{3}+2l)\Phi^e_k-\hat\zeta^e_k|_{B+}~, \nonumber\\
&\hat{F}_k \, = \, (\frac{4}{3}+2l)\Phi^e_k+\hat\zeta^e_k|_{B+}~.
\end{eqnarray}
Note again that it is justified to apply the matching conditions
since the universe is expanding on both sides of the matching
surface and thus the background also satisfies the matching
conditions.

By combining Eqs. (\ref{match1}) and (\ref{match2}) , we can
establish the relation between the gravitational potentials in
contracting and expanding phases. Since $\bar S_+$ is a decaying
mode, we will not write down its expression and focus our
attention instead on the dominant mode $\bar D_+$. In terms of the
modes in the contracting phase, it is given by
\begin{widetext}
\begin{eqnarray}\label{Dp}
\bar D_+ \, = \, \frac{(5+3w_c)(1+w_e)}{(1+w_c)(5+3w_e)}\bar{D}_- + &&
\frac{3(1+w_e)}{(5+3w_e)}c_s^2k^2 \times \bigg\{
\frac{\eta_{B+}-\eta_B}{\eta_{B-}-\eta_B}M_+(\frac{2\bar{D}_-}{3(1+w_c)}-\frac{\bar{S}_-}{(\eta_{B-}-\tilde\eta_{B-})^{2\nu_c}})
\nonumber\\ && \, -\frac{5+3w_c}{3(1+w_c)}M_+\bar{D}_- +
M_-(\bar{D}_-+\frac{\bar{S}_-}{(\eta_{B-}-\tilde\eta_{B-})^{2\nu_c}})
\bigg\}+O(k^4)~,
\end{eqnarray}
\end{widetext}
where we defined the parameters
\begin{eqnarray} \label{Mvalue}
M_{\pm} \, \equiv \, \frac{2}{9{\cal H}_{B\pm}^2(1+w^e_c)}+\frac{1}{y}~,
\end{eqnarray}
which are independent of $k$.

From the above result we see that both the constant and growing
modes of gravitational potential in the contracting phase effect
the dominant mode after the bounce. However, the growing mode is
suppressed by $k^2$ on large scales whereas the constant one
transfers through the bounce without a change in the spectral
index. These results agree with what is obtained using the
matching conditions at a singular hypersurface between the
contracting and the expanding phase, as shown in \cite{FF1}.

Inspecting our result (\ref{Dp}), we see that there are two ways
to obtain a scale-invariant spectrum of cosmological perturbations
after the bounce. The first is to consider a model in which the
$D-$ mode in the contracting phase has a scale invariant spectrum,
i.e. $D-(k) \sim k^{-3/2}$, the other is to take a scenario where
$S-(k) \sim k^{-7/2}$. As follows from (\ref{Dresult}, the first
possibility is realized if $p = \infty$, i.e. in an inflationary
contracting phase. The second way is realized in the case of a
matter-dominated contraction, a possibility already pointed out in
\cite{FB2} (see also \cite{Allen}). The Lee-Wick model yields a
natural realization of this way.

Let us now come back to our Lee-Wick background, and assume
quantum vacuum fluctuations. We insert the values $w_c=w_e=0$,
$c_s=1$ into (\ref{Dp}) and assume a symmetric fast bounce. Thus
\be \label{Dplw}
\bar{D}_+ \, = \, \bar{D}_-+\bigg[
-\frac{4}{5}\bar{D}_-+\frac{3}{5}\frac{{\cal
H}_{B-}^5}{2^4}\bar{S}_- \bigg]\frac{2k^2}{9{\cal H}_{B-}^2}~.
\ee
As discussed in the subsection on vacuum initial conditions, then
if the initial conditions are imposed at a time $\eta$
sufficiently early compared to the transition point $B-$, we have
$\Phi_k^{ini}\propto k^{-\frac{3}{2}}$ and therefore obtain $\bar
D_-\propto k^{\frac{3}{2}}$ and $\bar S_-\propto
k^{-\frac{7}{2}}$. Substituting these relations into Eq.
(\ref{Dplw}), one can see that whereas the contribution of $\bar
D_-$ to the final spectrum of $D+$ vanishes on large scales, the
contribution of $\bar S_-$ which starts out deep red is
blue-tilted by exactly the right amount to yield a final spectrum
proportional to $k^{-\frac{3}{2}}$ which is the scale-invariant
form.

To compute the amplitude of the spectrum, we insert the values
(\ref{Mvalue}) and the expression (\ref{Sspectrum}) for the value
of $S_-$ into (\ref{Dplw}) and use the background Friedmann
equation to replace the Hubble parameter by the energy density.
This yields
\be
\bar{D}_+ \, = \,  -\frac{\sqrt{\rho_{B-}}}{10\sqrt{2}M_p^2}
k^{-\frac{3}{2}}~.
\ee
Therefore, the power spectrum of the gravitational potential in
this case can be expressed as
\be \label{scalarresult}
P_{\Phi} \, \equiv \, \frac{k^3}{2\pi^2}|\bar{D}_+| ^2\, = \, \frac{\rho_{B-}}{(20\pi)^2M_p^4}~.
\ee

Note that as long as $M \ll M_{pl}$, the power spectrum of metric
fluctuations remains much smaller than $1$ and thus linear
perturbation theory is applicable throughout the bouncing phase.

\subsection{Numerical analysis}

Our analytical calculations involve approximations. Specifically,
in the contracting phase the scalar fields and hence the equation
of state are oscillating. But in our analytical analysis we have
replaced the time-dependent equation of state parameter by its
temporal average. It is thus important to confirm the results by
numerical integration of the full equations, namely Eq.
(\ref{Phieq}) coupled to the equation for the scalar matter field
fluctuation.

At first sight, it appears that the equation (\ref{Phieq})
contains a singularity at all turnaround points of $\phi$. Such
singularities are known from the study of the evolution of $\Phi$
during reheating taking into account the oscillatory nature of the
inflaton field \cite{FFRB,Lin}. However, this singularity is
actually not present. Let us consider in addition to the dynamical
perturbed Einstein equation (\ref{Phieq}) the perturbed Einstein
constraint equation
\be \label{pertconstr}
\Phi^{\prime} + {\cal H} \Phi \, = \, 4 \pi G \bigl( \phi^{\prime} \delta \phi
- \tphi^{\prime} \delta \tphi \bigr) \, .
\ee
Inserting (\ref{pertconstr}) into (\ref{Phieq}) yields
\bea \label{NewPhieq}
\Phi^{\prime \prime} &+& 6 {\cal H} {\Phi}^{\prime} +
2 \bigl( {\cal H}^{\prime} + 2 {\cal H}^2 \bigr) \Phi + k^2 \Phi  \nonumber \\
&=& \, 8 \pi G \bigl( 2 {\cal H} + \frac{\phi^{\prime \prime}}{\phi^{\prime}} \bigr)
 \phi^{\prime} \delta \phi \, ,
 \eea
from which it is clear that the singularity has disappeared. Thus,
we numerically solve (\ref{NewPhieq}) coupled to the perturbed
$\phi$ equation
 \be
 \delta \phi^{\prime \prime} + 2 {\cal H} \delta \phi^{\prime} +
 \bigl( k^2 + a^2 V_{\phi \phi} \bigr) \delta \phi \, = \, 4 \phi^{\prime} \Phi^{\prime} - 2 a^2 V_{\phi} \Phi
 \, ,
 \ee
where the subscripts on $V$ indicate the variables with respect to
which the potential is differentiated.

\begin{figure}[htbp]
\includegraphics[scale=0.8]{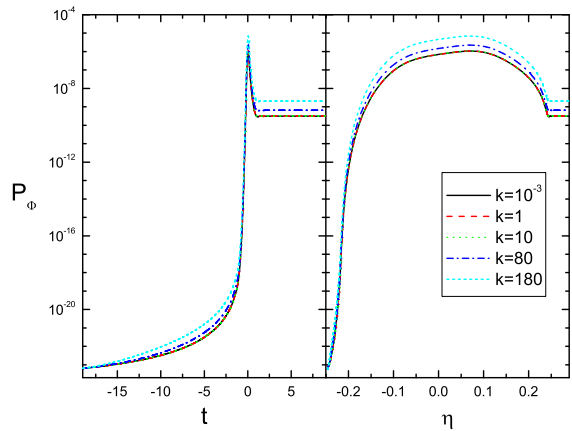}
\caption{Result of the numerical evolution of the curvature
perturbations with different comoving wavenumbers $k$ in the
Lee-Wick bounce. The horizontal axis in the left panel is cosmic
time, and in the right panel it is comoving time. The initial
values of the background parameters are the same as in Figure 1.
The units of the time axis are $M_{rec}^{-1}$, the comoving
wavenumber $k$ is unity for $k = M_{rec}$, as in Figure
\ref{bgfig}. } \label{fig:power1}
\end{figure}

\begin{figure}[htbp]
\includegraphics[scale=0.8]{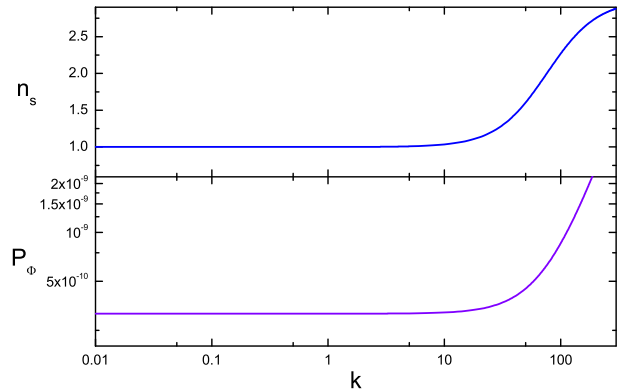}
\caption{Plot of the power spectrum of the curvature perturbation
$\Phi$ (lower panel) and of the spectral index (upper panel) as
functions of comoving wavenumbers $k$ in the Lee-Wick bounce. The
initial values of the background parameters are the same as in
Figure 1.} \label{fig:power2}
\end{figure}

{F}igures \ref{fig:power1} and \ref{fig:power2} show the results
of our numerical integration. The first figure shows the evolution
in time of the metric fluctuation $\Phi$ as a function of physical
time (left side) and conformal time (right side) for different
values of the comoving wavenumber $k$. We have chosen the bounce
point to correspond to physical and conformal time $0$. The
initial conditions for $\Phi$ were set at the initial time of the
simulation according to the vacuum initial condition prescription
discussed earlier. We see from this figure that before the bounce
the perturbations are dominated by the growing mode. When the
universe enters the bouncing phase, we see that the amplitude
approaches a constant and passes smoothly through the bounce. The
numerical evolution agrees well with the analytical solution
(\ref{asolb}) in the bouncing phase. After the bounce, the
perturbations are dominated by the constant mode. The numerical
evolution demonstrates that this mode can be inherited from the
growing mode in contracting phase.

{F}igure \ref{fig:power2} shows the  power spectrum of $\Phi$
(lower panel) and the spectral index $n_s$ (upper panel) as a
function of comoving wavenumber $k$.  On large scales (small
values of $k$), the power spectrum tends to a constant. The rise
of the spectrum for large values of $k$ is on scales which are
comparable to maximal value of the Hubble rate, i.e. for modes
which have not spent time outside of the Hubble radius.

\section{Gravitational Waves}

Now we turn to consider the evolution of gravitational waves
(tensor perturbations) in our background, assuming they start out
in the vacuum state on sub-Hubble scales in the contracting phase.
Since at the level of linear perturbation theory scalar metric
fluctuations and gravitational waves decouple, we can focus on a
metric containing only gravitational waves propagating in the
background. The standard form of this metric in a spatially flat
FRW background is
\be
ds^2 \, = \, a(\eta)^2[-d\eta^2 + (\delta_{ij}+\bar h_{ij})dx^idx^j]~,
\ee
where the Latin indexes run over the spatial coordinates, and the
tensor perturbation $\bar h_{ij}$ is real, transverse and
traceless, i.e.
\be
\bar h_{ij}  \, = \, \bar h_{ji}~;~~\bar h_{ii} \, = \, 0~;~~\bar h_{ij,j} \, = \, 0~.
\ee
Due to these constraints, we only have two degrees of freedom in
$\bar h_{ij}$ which correspond to two polarizations of
gravitational waves. For each polarization state (labelled by r in
the following), we can write $\bar h_{ij}(\eta, {\bf x})$ as a
scalar field $h^r(\eta, {\bf x})$ multiplied by a polarization
tensor $e^r_{ij}$ which is constant in space and time.

If matter contains an anisotropic stress tensor $\sigma_{ij}$,
there is a non-vanishing source term in the equation of motion for
tensor perturbations, namely
\be
\bar h_{ij}''+2\frac{a'}{a}\bar h_{ij}'-\nabla^2\bar h_{ij} \, = \, 16\pi
Ga^2\sigma_{ij}~.
\ee
If matter consists of a set of canonically normalized scalar
fields or a set of perfect fluids, there is no anisotropic stress
and thus no source term at linear order in perturbation theory for
gravitational waves.

As usual, we go to Fourier space The Fourier transformations of
the tensor perturbations and anisotropic stress tensor are give
by,
\be
\bar h_{ij}(\eta, {\bf x}) \, = \, 
\sum_{r = 1}^2 \int\frac{d^3k}{(2\pi)^\frac{3}{2}}h^r(\eta, {\bf
k})e^r_{ij}e^{i{\bf k}{\bf x}}~.
\ee
In order to canonically quantize the gravitational waves, it is
important to identify the variable in terms of which the action
has canonical kinetic term. This variable turns out to be (see
\cite{Mukhanov} for a derivation)
\be \label{vdef} v^r_k \, = \, \sqrt{\frac{(e^r)^i_j (e^r)^j_i}{32
\pi G}} a h^r_k \,
\ee
(where $h^r_k$ is a short hand notation for $h^r(\eta, {\bf k})$)
in terms of which the Einstein action expanded to second order in
$v^r$ becomes
\be
S \, = \, \sum_{r = 1}^r \frac{1}{2} \int \bigl( |(v^r)'_k|^2 - (k^2 - \frac{a''}{a})|v^r_k|^2 d\eta d^3k \, .
\ee
The resulting equation of motion for $v^r$ is
\be\label{eomt}
(v^r)'' + ( k^2 - \frac{a''}{a})v^r \, = \, 0~.
\ee

We are interested in computing the power spectrum of the tensor
modes. Making use of (\ref{vdef}), it is related to the power
spectrum of $v$ (which is the same for each polarization state)
via
\bea \label{tensorpower}
P^T_k(h) \, &=& \, a^{-2} 32 \pi G \sum_{r = 1}^2 P_k(v^r) \nonumber \\
&=& \, a^{-2} 64 \pi G \frac{k^3}{2 \pi^2} |v_k|^2 \, .
\eea
The tensor spectral index $n_T$ is defined by
\be
n_T \, \equiv \, \frac{d\,{\rm ln}\,P^T}{d\,{\rm ln}\,k}~.
\ee

The evolution of tensor perturbations is very similar to that of
scalar perturbations. Initially the perturbations are inside the
Hubble radius in the far past. Since the Hubble radius shrinks in
the contracting phase, the modes with small comoving wave number
exit the Hubble radius. After that the universe bounces to an
expanding phase, so these Fourier modes will return into the
Hubble radius.

In the current paper we focus on a mode with small $k$ so that it
exits the Hubble radius in the contracting phase (rather than the
bounce phase), then passes through the bounce and finally
re-enters the Hubble radius during the expanding phase.

We divide the time interval into three periods like we did for the
analysis of scalar metric fluctuations. During the phase when the
universe is contracting with an equation of state oscillating
around $w=0$, we have
\begin{eqnarray}
v \, &=& \, (\eta -\tilde\eta_{B-})^{\frac{1}{2}} \bigg\{
A^T_kJ_{-\frac{3}{2}}[k(\eta-\tilde\eta_{B-})] \nonumber\\
&& \, +B^T_kJ_{\frac{3}{2}}[k(\eta-\tilde\eta_{B-})]
\bigg\}~,
\end{eqnarray}
where $\tilde\eta_{B-} = \eta_{B-}-2/{{\cal H}_{B-}}$. Here, the
parameters $A^T_k$ and $B^T_k$ can be determined by the initial
condition for gravitational waves, which is  taken as the
Bunch-Davies vacuum
\be
v \, \sim \, e^{-ik\eta}/{\sqrt{2k}} \, .
\ee
So we have
\bea
A^T_k \, &=& \, i\frac{\sqrt{\pi}}{2} \,\,\, {\rm and} \nonumber \\
B^T_k \, &=& \, -\frac{\sqrt{\pi}}{2} \, .
\eea
Therefore, the asymptotic form of the solution to the tensor
perturbation in the contracting phase is
\begin{eqnarray}
  v(k,\eta) \, = \, \left\{ \begin{array}{c}
    -\frac{i}{\sqrt{2}}k^{-\frac{3}{2}}(\eta-\tilde\eta_{B-})^{-1},~~{\rm outside~Hubble~radius}; \\
    \\
    \frac{1}{\sqrt{2k}}e^{-ik(\eta-\tilde\eta_{B-})},~~{\rm
    inside~Hubble~radius} .
\end{array} \right.  \label{v1}
\end{eqnarray}

During the bouncing phase, we have the approximate relation
\be
\frac{a''}{a} \, \simeq \, \frac{4}{\pi}\alpha a_B^2 \, = \, \frac{y}{3} \, .
\ee
Solving Eq. (\ref{eomt}), we have
\begin{eqnarray}
  &&v(k,\eta)= \nonumber\\ &&\left\{ \begin{array}{c}
    C^T_k\cos[l(\eta-\eta_B)]+D^T_k\sin[l(\eta-\eta_B)],~~{\rm k^2\geq\frac{y}{3}}; \\
    \\
    C^T_ke^{l(\eta-\eta_B)}+D^T_ke^{-l(\eta-\eta_B)},~~{\rm
    k^2<\frac{y}{3}},
\end{array} \right.  \label{v2}
\end{eqnarray}
where we define $l^2 = |k^2-\frac{y}{3}|$. Since the Hubble
parameter approaches zero when the universe is transiting from the
contracting to the expanding phase, all fluctuation modes return
to the sub-Hubble region, but only for a very brief time. However,
from the above solution we interestingly find that $k_{ph}^2(\sim
k^2/a_B^2)$ and $\dot H(\sim\alpha)$ are comparable.

After the bounce, an expanding phase with its EoS $w=0$ takes
place. So the solution to the gravitational waves is given by
\begin{eqnarray}
v \, &=& \, (\eta-\tilde\eta_{B+})^{\frac{1}{2}} \times \bigg\{
E^T_kJ_{-\frac{3}{2}}[k(\eta-\tilde\eta_{B+})] \nonumber\\
&&+F^T_kJ_{\frac{3}{2}}[k(\eta-\tilde\eta_{B+})]
\bigg\}~,\label{v3}
\end{eqnarray}
where $\tilde\tau_{B+}=\tau_{B+}-2/{{\cal H}_{B+}}$. This solution
takes on the asymptotic form,
\begin{eqnarray}\label{v3a}
v \, \simeq \,
\sqrt{\frac{2}{\pi}}\frac{F^T_k}{3}k^{\frac{3}{2}}(\eta-\tilde\eta_{B+})^2~,
\end{eqnarray}
when the modes are outside the Hubble radius.

Having obtained the solutions of the tensor perturbations in the
different phases, now we need to match these solutions and
determine the coefficients $C^T_k$, $D^T_k$, $E^T_k$ and $F^T_k$
respectively. This procedure is analogous to the matching process
of scalar perturbations performed in the previous section. For a
non-singular bounce scenario such as the bounce we are
considering, the continuity of the background evolution implies
that both $v$ and $v'$ are able to pass through the bounce
smoothly. So we match $v$ and $v'$ in (\ref{v1}) and (\ref{v2}) on
the surface $\tau_{B-}$, and those in (\ref{v2}) and (\ref{v3}) on
the surface $\tau_{B+}$. With these matching conditions, we can
determine all the coefficients and finally get the solution for
$v$ at late times.

Since in the specific model we considered in this paper, the
evolution of the universe is symmetric with respective to the
bounce point, we can simply take ${\cal H}_{B-}\simeq-{\cal
H}_{B+}$. In addition, we have shown that the bounce takes place
very fast on the time scale set by $k^{-1}$, so we have
$l(\tau_{B+}-\tau_{B-})\gg1$. Therefore, we eventually obtain the
approximate result
\be
F^T_k  \, \simeq \,  i\frac{\sqrt{\pi}{\cal H}_{B+}^3}{8k^3}
\ee
and the asymptotical form of $v$ in the final stage can be expressed as
\be \label{vf}
v^f \rightarrow i\frac{\sqrt{2}}{24} \frac{{\cal
H}_{B+}^3}{k^{\frac{3}{2}}} (\eta-\tilde\eta_{B+})^{2}~.
\ee

Now we are able to derive the power spectrum of primordial
gravitational waves. From the definition of Eq. (\ref{tensorpower}),
the primordial power spectrum is given by
\begin{eqnarray}\label{PT}
P_T (k) \, &=& \, G \frac{32 k^3}{\pi} |\frac{v^f}{a}|^2\nonumber\\
    &=& \, \frac{2\rho_{B+}}{27\pi^2M_p^4} ~.
\end{eqnarray}
From Eq. (\ref{PT}), we can read that the spectrum are
scale-invariant on large scales (which is consistent with the
result in Ref. \cite{Cai:2008ed}).

Comparing our result of the tensor power spectrum with the result
(\ref{scalarresult}) for the power spectrum of scalar metric
fluctuations, we obtain a tensor to scalar ratio of the order of
$30$, which is in excess of the current observational bounds. The
exact value of the ratio, however, will depend on the detailed
modelling of the bounce phase \footnote{We have modelled the
contracting and expanding phases with a background with a constant
equation of state. In this case, the squeezing factor for scalar
and tensor cosmological perturbations is the same. If the equation
of state changes, then the amplitude of the scalar spectrum can be
enhanced, as happens in inflationary cosmology at the time of
reheating \cite{BST,BK}. Since $1 + w$ does not change by orders
of magnitude in the Lee-Wick bouncing cosmology - whereas it does
change by orders of magnitude in the case of inflationary
cosmology - we do not expect a dramatic enhancement of the scalar
power spectrum. However, we do expect some increase in the
amplitude. A detailed study of this issue is left to further
work.} However, the conclusion that the tensor to scalar ratio
will be rather large will be robust, and also agrees with the
analysis of \cite{Allen} done in a different context.

\section{Conclusions and Discussion}

Recently, the Lee-Wick Standard Model has been suggested as an
extension of the Standard Model of particle physics providing an
alternative to supersymmetry in terms of addressing the hierarchy
problems.

In this paper, we have considered the cosmology of the Higgs
sector of the Lee-Wick Standard Model, an alternative to
supersymmetry to solving the hierarchy problem. We have found that
homogeneous and isotropic solutions are non-singular. Thus, the
Lee-Wick model provides a possible solution of the cosmological
singularity problem.

We then considered the spectrum of cosmological perturbations and
find that quantum vacuum fluctuations established in the
contracting phase evolve into a scale-invariant spectrum in the
expanding phase. Note that these results emerge without having to
introduce any additional features into the model, unlike the
situation in inflationary cosmology where the existence of a new
scalar field satisfying slow-roll conditions must be assumed, or
the situation in Ekpyrotic models where once again a scalar field
with special features must be assumed.

Tuning the amplitude of the spectrum of scalar metric fluctuations
to agree with the amplitude inferred from CMB observations
\cite{COBE}, we can determine the scale $M$ of the new physics
which is present in the Lee-Wick model. The required value of $M$
turns out to be about $10^{17}$GeV.

We have also computed the spectrum of gravitational waves and also
find a scale-invariant spectrum assuming that the fluctuations are
quantum vacuum in nature. The tensor to scalar ratio may be in
excess of the current observational bounces, but the exact value
will depend on the detailed modelling of the bounce phase.

One of the main successes of cosmological inflation is the
solution of the horizon, homogeneity, size and flatness problems
of Standard Big Bang cosmology which it provides. How does a
bouncing cosmology such as our Lee-Wick model measure up against
these successes? First of all, if the universe starts out large
and cold, there are no horizon and size problems. If the spatial
curvature at temperatures in the contracting phase comparable to
the current temperature is not larger than the current spatial
curvature, then there will be no flatness problem either because
the deviation of $\Omega_K$ from 0 decreases in the contracting
phase at the same rate that it increases in the expanding phase.
The key challenge for any bouncing cosmology is to control the
magnitude of the inhomogeneities and to provide a mechanism for
preventing the universe to collapse into a gas of black holes at
the end of the phase of contraction. For an attempt to address
this issue in the case of string gas cosmology see \cite{Nima}.

We would like to conclude this paper by putting our work in the
context of previous work on perturbations in bouncing cosmologies.
The issue of the mixing of the $S$ and $D$ modes at a cosmological
bounce has been hotly debated in the literature since the
Ekpyrotic scenario was proposed. In the case of the Ekpyrotic
scenario, for vacuum initial conditions the $S$ mode of $\Phi$
inherits a scale-invariant spectrum, whereas the $D$ mode obtains
a blue spectrum with index $n = 3$ \cite{Lyth,Khoury,FF1,Hwang}
(see also the more recent analysis of \cite{Creminelli} and the
recent review of \cite{Wands2}). This is also the spectrum of
$\zeta$. According to the Hwang-Vishniac \cite{HV}
(Deruelle-Mukhanov \cite{DM}) matching conditions applied at a
hypersurface on which we glue the expanding to the contracting
universe, the mixing between the $S-$ mode and the $D+$ mode is
suppressed by a power of $k^2$ (see e.g. \cite{FF1} for a
discussion of this point). Hence, the spectrum of metric
fluctuations after the bounce is not scale-invariant. The
Pre-Big-Bang scenario faces a similar problem \cite{BGGMV}. These
conclusions were confirmed in some specific models in which the
bounce was smoothed out by making use of higher derivative gravity
terms (see \cite{Copeland} in the case of the Pre-Big-Bang model
and \cite{Tsujikawa,Cardoso} in the case of the Ekpyrotic
scenario). However, the use of the matching conditions was
challenged in \cite{Durrer} where it was pointed out that if the
background solution does not satisfy the matching conditions at
the bounce, there is no reason to expect the fluctuations to do
so. In fact, in the case of the Ekpyrotic scenario (which is
intrinsically a higher-dimensional cosmology), computations done
in the higher dimensional framework yielded a successful transfer
of the scale-invariant spectrum of metric fluctuations from the
contracting to the expanding phase \cite{Tolley}, a conclusion
which was confirmed in \cite{McFadden} and, in a slightly
different setting, in \cite{BBP} \footnote{It wold go beyond the
scope of this discussion to review attempts to preserve a
scale-invariant spectrum in Pre-Big-Bang and Ekpyrotic scenarios
via introducing entropy fluctuations.}.

Calculations have also been done in some other non-singular
bouncing models \cite{Peter}. For example, studies done in models
in which the bounce is induced by a negative energy density scalar
field found no unsuppressed matching between the growing
perturbation mode in the contracting phase and the constant mode
in the expanding phase \cite{Bozza,FabioPeter}, in contrast to
what was obtained in some initial work \cite{Fabio,Peter1}. Both
studies in models in which a bounce was generated by a curvature
term in the Einstein action \cite{Peter2} and analyses in some
other bouncing models \cite{Abramo,Peter4} yielded un-suppressed
matching of the dominant modes of the contracting and expanding
phases.

The upshot of these analyses is that the transfer of fluctuations
through a cosmological bounce can depend quite sensitively on the
physics of the bounce.

It was realized that the equation of motion for $\zeta$ has
singularities in the case of a non-singular bounce, thus casting
doubt on the belief that in all cases $\zeta$ is conserved at a
bounce. It was shown that the $\Phi$ equation is free of such
singularities and is thus a safer equation to use \cite{BFS,ABB}.
In our previous work \cite{CQBPZ} it was shown in the case of the
quintom bounce model that there is unsuppressed mixing between the
$D+$ and $S-$ modes on length scales which are small compared to
the duration of the bouncing period, whereas on longer length
scales the mixing is suppressed (but not completely absent). In
the present work, the bounce is short compared to the length
scales we are interested in.

Our work shows that the evolution of fluctuations through the
non-singular bounce in the Lee-Wick model is rather standard.
There is no un-suppressed coupling between the dominant modes of
the contracting and expanding phases, and $\zeta$ is conserved at
the bounce.

In the current paper we have not considered radiation. Since the
energy density in radiation increases faster than that in matter,
radiation would dominate at early times. However, in the Lee-Wick
standard model there is a Lee-Wick partner to each field. In
particular, there is a Lee-Wick photon partner $\tilde{\gamma}$ of
the radiation field $\gamma$. At high energy densities, then as a
consequence of interactions between $\gamma$ and $\tilde{\gamma}$,
we expect that energy will flow from $\gamma$ into
$\tilde{\gamma}$, like it flows from $\phi$ to $\tilde{\phi}$ in
our scalar field model. Then, a cosmological bounce would occur in
a manner similar to how it occurs in our model. Adding
intermediate phases of radiation between the bouncing phase and
the contracting and expanding matter phases will not change our
results concerning the spectrum of fluctuations for modes which
exit the Hubble radius during the phase of matter domination,
which are the modes we are interested in when trying to explain
the large-scale structure of the universe and the CMB
anisotropies. A study of these issues is left to a followup paper.

It would also be interesting to consider entropy fluctuations and
non-Gaussian signatures of our scenario. We leave these topics for
future research.

Note added: while this paper was being prepared for submission, a
preprint appeared \cite{SLee} pointing out that the Lee-Wick model
provides a realization of the quintom scenario and could be
applied to study the current acceleration of the universe. We find
it more natural to consider the corrections to the cosmological
evolution which are obtained in the very early universe.

\begin{acknowledgments}

We wish to thank Andy Cohen, Ben Grinstein, Hong Li, Jie Liu and
Mark Wise for useful discussions. RB wishes to thank the Theory
Division of the Institute of High Energy Physics (IHEP) for their
wonderful hospitality and financial support. RB is also supported
by an NSERC Discovery Grant and by the Canada Research Chairs
Program. The research of X.Z., Y.C. and T.Q. is supported in part
by the National Science Foundation of China under Grants No.
10533010 and  10675136, by the 973 program No. 2007CB815401, and
by the Chinese Academy of Sciences under Grant No. KJCX3-SYW-N2

\end{acknowledgments}

\end{document}